\documentclass[12pt]{article}

\begin{document}

\begin{center}

{\Large {Quark mixing angles vs quark masses: potential approach}}

\vspace{1,5cm}

{Boris L. Altshuler}\footnote{E-mail addresses: baltshuler@yandex.ru \,\,\,  \&  \,\,\,  altshul@lpi.ru}

\vspace{1cm}

{\it {Theoretical Physics Department, P.N. Lebedev Physical
Institute, \\  53 Leninsky Prospect, Moscow, 119991, Russia}}

\vspace{1,5cm}

\end{center}

{\bf Abstract:} It is shown that following phenomenologically favorable expressions of quark mixing angles through the ratios of current quark masses: $\sin\theta_{12} = \sqrt{m_{d} / |m_{s}|}$, $\sin\theta_{23} = 2 \, |m_{s}| / m_{b}$, $\sin\theta_{13} = \sqrt{m_{d} |m_{s}|} / m_{b}$ may be derived as stable points of certain 4-th power in $V_{CKM}$ flavor-invariant potentials built with traces of 3x3 quark up and down mass matrices.

\vspace{0,5cm}


\vspace{0,5cm}


\newpage

\section{Preliminaries and motivation}

\quad The arbitrariness of fermion masses, mixing angles and CP violating phases of the quark and lepton sectors of Standard Model (SM) plague the SM for decades in spite of the plural efforts to find theory explaining these fundamental numbers, see \cite{Isidori} - \cite{Feruglio} and references therein. Here the quark sector of SM will be considered which fermion Lagrangian $L_{f}$ \cite{Peskin} we supplement with potential $W(\hat{M}_{u}, \hat{M}_{d})$ depending on 3x3 up and down current quark mass matrices:

\begin{equation}
\label{1}
L_{f} = \bar{\psi}_{u}\hat{\partial}\psi_{u} + \bar{\psi}_{d}\hat{\partial}\psi_{d} + \psi^{*}_{uL}\hat{M}_{u}\psi_{uR} + \psi^{*}_{dL}\hat{M}_{d}\psi_{dR} - W(\hat{M}_{u}, \hat{M}_{d}).
\end{equation}
Potential $W$ will be discussed in Sec. 2. $\hat{M}_{u, d} = v \hat{Y}_{u,d}$ ($v$ is vacuum average of the Higgs field, $\hat{Y}_{u,d}$ are dimensionless matrices which elements are so called Yukawa couplings). In Eq. (\ref{1}) SM gauge fields are omitted, and also are omitted generation (flavor) indices of spinor fields ($\psi_{u} = u, c, t$ quarks; $\psi_{d} = d, s, b$ quarks) and of their mass matrices. It is well known \cite{Peskin} that with proper unitary transformations of the left and right Fermi fields $\hat{M}_{u,d}$ may be made hermitian and diagonalized: $\hat{M}_{u} = U_{u}\hat{m}_{u}U^{\dag}_{u}$, $\hat{M}_{d} = U_{d}\hat{m}_{d}U^{\dag}_{d}$; $\hat{m}_{u} = {\rm{diag}}(m_{ui}) = {\rm{diag}}(m_{u}, m_{c}, m_{t})$; $\hat{m}_{d} = {\rm{diag}}(m_{dj}) = {\rm{diag}}(m_{d}, m_{s}, m_{b})$; $i,j = 1,2,3$; $m_{f}$ ($f = u, c, t, d, s, b$) are corresponding six current quark masses; these mass eigenvalues may be positive or negative. In case matrices $\hat{M}_{u,d}$ can't be diagonalized with one and the same unitary transformation the vertex of quarks interaction with charged gauge bosons $W^{\pm}$ includes, in the SM mass eigenvalues basis, the flavor-mixing unitary Cabibbo-Kobayashi-Maskawa matrix $\hat{V}_{CKM} = U^{\dag}_{u}U_{d}$ \cite{Peskin}. In standard PDG parametrization $\hat{V}_{CKM}$ ($\equiv \hat{V}$ below) is given by three mixing angles $\theta_{12}$, $\theta_{23}$, $\theta_{13}$ and the CP-violating phase $\delta$, and looks as:

\begin{equation}
\label{2}
\hat{V} = \left(\begin{array}{ccc}
1 & 0 & 0 \\
0 & c_{23} & s_{23} \\
0 & -s_{23} & c_{23}
\end{array}\right) \times
\left(\begin{array}{ccc}
c_{13} & 0 & s_{13} e^{-i\delta} \\
0 & 1 & 0 \\
- s_{13} e^{i\delta} & 0 & c_{13}
\end{array}\right) \times
\left(\begin{array}{ccc}
c_{12} & s_{12} & 0 \\
-s_{12} & c_{12} & 0 \\
0 & 0 & 1
\end{array}\right),
\end{equation}
here $s_{12} = \sin\theta_{12}$, $c_{12} = \cos\theta_{12}$ etc.

Thus two ermitian matrices $\hat{M}_{u,d}$ hide in ten experimentally observed constants of the SM quark sector: masses of six quarks and four named above parameters of CKM matrix. The updated values of the current quarks masses (at scale 2 GeV; errors are shown in brackets) are presented in \cite{masses}: $m_{u}$ = 2.16(38) MeV, $m_{c}$ = 1.27(2) GeV, $m_{t}$ = 172.4(7) GeV; $m_{d}$ = 4.67(32) MeV, $m_{s}$ = 0.093(8) GeV, $m_{b}$ = 4.18(2) GeV (if the eigenvalues of matrices $\hat{M}_{u,d}$ are negative, then these numbers refer to the modules of the eigenvalues). CKM-angles are also observed with high accuracy \cite{angles}.

The observed values of these constants obey hierarchical structure. In particular, there is the so far unexplained correlation (which is the main focus of the present paper) of hierarchy of quark masses and hierarchy of the CKM matrix mixing angles. These hierarchies and their correlation are immediately visible if the CKM angles and ratios of quark masses are expressed with one and the same Wolfenstein parameter $\lambda$ \cite{Wolfenstein}:

\begin{equation}
\label{3}
\begin{array}{c}
\lambda = s_{12} = 0.2253(7); \, \, \, \, \, s_{23} = 4.080(14)\cdot 10^{-2} = 0.81 \, \lambda^{2}; \\
s_{13} = 3.82(20) \cdot 10^{-3} = 0.34 \, \lambda^{3}; \\
\frac{m_{u}}{m_{c}} = 1.70(40) \cdot 10^{-3} = 0.65 \, \lambda^{4}; \qquad \frac{m_{d}}{m_{s}} = 0.050(7) = \lambda^{2}; \qquad \qquad \quad \\
\frac{m_{c}}{m_{t}} = 0.747(12) \cdot 10^{-2} = 2.9 \, \lambda^{4}; \qquad \frac{m_{s}}{m_{b}} = 2.22(25) \cdot 10^{-2} = 0.44 \, \lambda^{2}; \, \, \\
\frac{m_{u}}{m_{t}} = 1.26(20) \cdot 10^{-5} = 1.88 \, \lambda^{8}; \qquad \frac{m_{d}}{m_{b}} = 1.12(8) \cdot 10^{-3} = 0.44 \, \lambda^{4}; \, \, \, \, \,
\end{array}
\end{equation}
the values of quark masses given above are used here; values of sinuses of the CKM angles see in \cite{angles}.

The task to explain the origin of these numbers and their specific tuning is called "flavor puzzle". The part of this "puzzle" is to find the well grounded expressions for the CKM mixing angles through the given in (\ref{3}) ratios of quark masses. There were many attempts to do it. Among them may be called "texture zeroes" approach where, because of some natural symmetry conditions imposed upon the primary $\hat{M}_{u,d}$ matrices, some of their elements prove to be zeros, hence bi-unitary transformation described above give formulas for the mixing angles more or less compatible with values given in (\ref{3}), see \cite{Xing}, \cite{Wilczek}, \cite{Fritzsch}  and references therein. But we shall go here another way.

\section{Potential approach: calculation of mixing angles}

\quad The idea that Yukawa couplings are dynamical fields and that their observed values are these fields' vacuum expectation values to be determined by an extremizing of some flavor-invariant potential looks quite attractive. Actually, this idea goes back to N. Cabibbo \cite{Cabibbo}. Spontaneous breaking of a flavor symmetry may be a natural explanation of the observed flavor structure. The unseen Goldstone bosons and other difficulties of this approach and the possible ways to overcome them are discussed in literature, see \cite{Isidori}(a), \cite{Feruglio}(a), \cite{Alonso1} - \cite{difficulties} and references therein, where both - SM quark and lepton sectors are considered. We shall not touch these issues here, but just try to build potential introduced in (\ref{1}) which extremizing over mixing angles must give experimentally viable dependencies of these angles on the observed quarks mass ratios (\ref{3}).

Jarlskog determinant of the commutator of $\hat{M}_{u}$ and $\hat{M}_{d}$ matrices  is the most famous flavor-invariant \cite{Jarlskog}; its non-zero value signals the CP violation. This determinant may be expressed through certain trace of mass matrices \cite{Branco}:

\begin{equation}
\label{4}
\begin{array}{c} 
{\rm{Det}}\{[\hat{M}_{u},\hat{M}_{d}]\} = \frac{1}{3}\,{\rm{Tr}}\{[\hat{M}_{u},\hat{M}_{d}]^{3}\} = 2i \, {\rm{Im}} {\rm{Tr}}\{\hat{M}_{u}\hat{M}_{d}\hat{M}_{u}^{2}\hat{M}_{d}^{2}\} = \\
(m_{c} - m_{u})\,(m_{t} - m_{u})\,(m_{t} - m_{c})\, (m_{s} - m_{d})\,(m_{b} - m_{d})\,(m_{b} - m_{s})\,2 i J; \\
J = s_{12}c_{12}s_{23}c_{23}s_{13}c_{13}^{2}\, \sin\delta = 3.038 \cdot 10^{-5}.
\end{array}
\end{equation}
Last figure is an experimentally observed value of $J$. In what follows we shall use real part of ${\rm{Tr}}\{\hat{M}_{u}\hat{M}_{d}\hat{M}_{u}^{2}\hat{M}_{d}^{2}\}$ (\ref{4}) as well as some other traces that include second and fourth powers of the CKM matrix. In \cite{invariants} all independent traces of two 3x3 matrices are listed which number is limited according to the Caylay-Hamilton theorem. In \cite{Bingrong1}, \cite{Lu} the flavor invariants in leptonic sector, which is essentially more complicated than the quark sector, are investigated. Papers \cite{Bingrong2} consider flavor invariants as the sources of CP violation in the canonical seesaw model for neutrino masses and that in the seesaw effective field theory; it is shown that all the physical parameters can be extracted using the primary invariants (see also \cite{Jarlskog} and references therein). This fact demonstrates in particular the flavor invariance of the results of the present paper, although these results are obtained in specific flavor basis and for specific parametrization (\ref{2}) of the CKM matrix. 

Developing the flavor-invariant approach of papers \cite{Jarlskog} - \cite{Bingrong2} the explicit dependences of some flavor invariants on quark masses and CKM matrix parameters are derived in the present paper; also the concrete number results for the values of mixing angles are obtained from the extremization of potential $W(\hat{M}_{u}, \hat{M}_{d})$ (\ref{1}) built of certain combinations of flavor invariants. The building blocks for potential $W(\hat{M}_{u}, \hat{M}_{d})$ are the flavor invariant traces ${\rm{Tr}}\{\hat{M}_{u}^{n_{1}}\} = \sum_{i=1}^{3}m_{ui}^{n_{1}}$, ${\rm{Tr}}\{\hat{M}_{d}^{n_{2}}\} = \sum_{j=1}^{3}m_{dj}^{n_{2}}$, and "mixing" traces of type: ${\rm{Tr}}\{\hat{M}_{u}^{n_{1}}\hat{M}_{d}^{n_{2}}\} = {\rm{Tr}}\{\hat{m}_{u}^{n_{1}}\hat{V}\hat{m}_{d}^{n_{2}}\hat{V}^{\dag}\}$, ${\rm{Tr}}\{\hat{M}_{u}^{n_{1}}\hat{M}_{d}^{n_{2}}\hat{M}_{u}^{n_{3}}\hat{M}_{d}^{n_{4}}\} = {\rm{Tr}}\{\hat{m}_{u}^{n_{1}}\hat{V}\hat{m}_{d}^{n_{2}}\hat{V}^{\dag}\hat{m}_{u}^{n_{3}}\hat{V}\hat{m}_{d}^{n_{4}}\hat{V}^{\dag}\}$. General expressions for these two "mixing" traces as functions of moduli of the "angle" elements of arbitrary unitary matrix $\hat{V}$ are given in Appendix for, also arbitrary, diagonal matrices that alternate with $\hat{V}$ and $\hat{V}^{\dag}$.

We shall consider potential $W$ in (\ref{1}) to be symmetric with exchange $\hat{M}_{u} \leftrightarrow \hat{M}_{d}$ and homogeneous in masses. The looked for mixing angles will depend on mass ratios (\ref{3}); more of that: because of strong inequality of the "up" and "down" mass ratios the main contributions in expressions for mixing angles come only from the "down" ratios $m_{d} / m_{s}$, $m_{s} / m_{b}$, $m_{d} / m_{b}$.

\vspace{0,5cm}

\quad {\large {\bf{2.1. Calculation of $s_{12}$.}}}

\vspace{0,5cm}
Let us look at the 4-th power in masses potential when $\theta_{23}$, $\theta_{13}$ in (\ref{2}) are zero and only $\theta_{12} \ne 0$:

\begin{equation}
\label{5}
\begin{array}{c}
W^{(4)} = {\rm{Re}} \, \large[{\rm{Tr}}\{\hat{M}_{u}\hat{M}_{d}\hat{M}_{u}\hat{M}_{d}\} - 2 \, {\rm{Tr}}\{\hat{M}_{u}^{2}\hat{M}_{d}^{2}\}\large] = \nonumber \\ 
= {\rm{Re}} \, \large[{\rm{Tr}}\{\hat{m}_{u}\hat{V}\hat{m}_{d}\hat{V}^{\dag}\hat{m}_{u}\hat{V}\hat{m}_{d}\hat{V}^{\dag}\}
- 2 \, {\rm{Tr}}\{\hat{m}_{u}^{2}\hat{V}\hat{m}_{d}^{2}\hat{V}^{\dag}\}\large] \approx \\
\approx - \sum_{i=1}^{3}m_{ui}^{2}m_{di}^{2} + m_{c}^{2} [m_{s}^{2}s_{12}^{4} + 2 \, m_{s}m_{d}s_{12}^{2}]; \\
\bar{s}_{12} = \sqrt{\frac{m_{d}}{|m_{s}|}} \, \, \, \, .
\end{array}
\end{equation}

The last approximate expression in the chain of equations (\ref{5}) is received from the general formulas (\ref{19}), (\ref{21}) of Appendix, where only contributions of the largest mass ratios were taken into account. It is seen that potential (\ref{5}) has stable nonzero minimum $s_{12} = \bar{s}_{12}$ for the negative sign of the second eigenvalue $m_{s}$ of mass matrix $\hat{M}_{d}$; whereas an unstable extremum of this potential, at $s_{12} = 0$, corresponds to the unbroken flavor symmetry $\hat{V}_{CKM} \equiv \hat{V} = \hat{1}$. Formula for Cabibbo angle $\sin\theta_{\rm{Cab}} = \bar{s}_{12} = \sqrt{m_{d}/|m_{s}|}$ is known for decades, but, as to our knowledge, it was not derived earlier from the extremization of potential of type (\ref{5}).

The analogous calculation of potential (\ref{5}) in case $s_{23} \ne 0$, $s_{12} = s_{13} = 0$ gives for $s_{23}$ the similar to (\ref{5}) 4-th power potential which flavor braking minimum at $\bar{s}_{23} = \sqrt{|m_{s}|/m_{b}}$ is in gross contradiction with experimental data (\ref{3}). Also nothing good will come out from attempt to calculate $s_{13}$ from potential (\ref{5}).

\newpage

\vspace{0,5cm}
 \quad {\large {\bf{2.2. Calculation of $s_{23}$, $s_{13}$.}}}

\vspace{0,5cm}

To get the sensible extremal values of $s_{23}$ and $s_{13}$ more complicated 8-th power in masses potential $W$ (\ref{1}) must be considered:

\begin{equation}
\label{6}
\begin{array}{c}
W = \kappa \, W^{(8)} = \kappa \, {\rm{Re}} \, \large[c_{1}W_{1} + c_{2} W_{2} + c_{3} W_{3} + c_{4} W_{4}\large], \\
W_{1} = ({\rm{Tr}}\hat{M}_{u})^{2} \cdot ({\rm{Tr}}\hat{M}_{d})^{2} \cdot [{\rm{Tr}}\{\hat{M}_{u}\hat{M}_{d}\hat{M}_{u}\hat{M}_{d}\} - 2 \, {\rm{Tr}}\{\hat{M}_{u}^{2}\hat{M}_{d}^{2}\}]; \nonumber \\
W_{2} = {\rm{Tr}}\hat{M}_{u} \cdot {\rm{Tr}}\hat{M}_{d} \cdot [{\rm{Tr}}\{\hat{M}_{u}\hat{M}_{d}\hat{M}_{u}^{2}\hat{M}_{d}^{2}\} - 2 \, {\rm{Tr}}\{\hat{M}_{u}^{3}\hat{M}_{d}^{3}\}]; \nonumber \\
W_{3} = {\rm{Tr}}\{\hat{M}_{u}\hat{M}_{d}\hat{M}_{u}^{3}\hat{M}_{d}^{3}\} - 2 \, {\rm{Tr}}\{\hat{M}_{u}^{4}\hat{M}_{d}^{4}\}; \nonumber \\
W_{4} = {\rm{Tr}}\{\hat{M}_{u}^{2}\hat{M}_{d}^{2}\hat{M}_{u}^{2}\hat{M}_{d}^{2}\} - 2 \, {\rm{Tr}}\{\hat{M}_{u}^{4}\hat{M}_{d}^{4}\}, \nonumber
\end{array}
\end{equation}
($\kappa$ is a dimensional constant irrelevant here).

With use of general formulas of Appendix and leaving only the largest terms in the hierarchy of mass ratios, we present the expressions for potential (\ref{6}) in three cases ($W^{(8)}_{0}$ is an unessential here value of potential (\ref{6}) at $\hat{V} = \hat{1}$):

1) $s_{12} \ne 0$, $s_{23} = s_{13} = 0$ in (\ref{2}):

\begin{equation}
\label{7}
W^{(8)}(s_{12}) \approx W^{(8)}_{0} + m_{t}^{2} \, m_{c}^{2} \, m_{b}^{2} \cdot c_{1} \, [m_{s}^{2} \, s_{12}^{4} - 2\,|m_{s}| \, m_{d} \, s_{12}^{2}].
\end{equation}
$c_{1} > 0$ is a necessary condition for potential (\ref{7}) to have minimum at $s_{12} = {\bar{s}}_{12}$ (see (\ref{5})).

2) $s_{23} \ne 0$, $s_{12} = s_{13} = 0$:

\begin{equation}
\label{8}
\begin{array}{c}
W^{(8)}(s_{23}) \approx W^{(8)}_{0} + m_{t}^{4} \, m_{b}^{2} \cdot \{s_{23}^{4} \, m_{b}^{2} \, \sum_{k=1}^{4}c_{k} - \\
- s_{23}^{2}\, [|m_{s}| \, m_{b} \, (2c_{1} + c_{2} + c_{3}) - 2 \, m_{s}^{2}\, (c_{1} + c_{2} + c_{4})]\}.
\end{array}
\end{equation}
There are three conditions ($\sum_{k=1}^{4}c_{k} > 0$; $2c_{1} + c_{2} + c_{3} = 0$; $c_{1} + c_{2} + c_{4} < 0$) that are necessary for potential (\ref{8}) to have stable minimum at

\begin{equation}
\label{9}
s_{23} = \bar{s}_{23} = \sqrt{\frac{|c_{1} + c_{2} + c_{4}|}{\sum_{k=1}^{4}c_{k}}} \cdot \frac{|m_{s}|}{m_{b}}.
\end{equation}

3) $s_{13} \ne 0$, $s_{12} = s_{23} = 0$:

\begin{equation}
\label{10}
\begin{array}{c}
W^{(8)}(s_{13}) \approx W^{(8)}_{0} + m_{t}^{4} \, m_{b}^{2} \cdot \{s_{13}^{4} \, m_{b}^{2} \, \sum_{k=1}^{4}c_{k} - \\
- s_{13}^{2}\, [- m_{d} \, m_{b} \, (2c_{1} + c_{2} + c_{3}) - 2 \, m_{d}^{2}\, (c_{1} + c_{2} + c_{4}) + m_{d}\, |m_{s}| \, (4c_{1} + c_{2})]\}.
\end{array}
\end{equation}
It can be seen that in addition to the conditions imposed on constants $c_{k}$ in item 2 above, one more condition, $4c_{1} + c_{2} > 0$, is necessary for potential (\ref{10}) to have a stable minimum at

\begin{equation}
\label{11}
s_{13} = \bar{s}_{13} = \sqrt{\frac{4c_{1} + c_{2}}{2 \, \sum_{k=1}^{4}c_{k}}} \cdot \frac{\sqrt{m_{d}\,|m_{s}|}}{m_{b}};
\end{equation}
this expression was obtained from (\ref{10}) by neglecting the penultimate term on the RHS of (\ref{10}), taking into account that $m_{d}^{2} \ll m_{d}\, |m_{s}|$.

The set of constants $c_{k}$ in (\ref{6}) with the smallest integer values that satisfy the above conditions: $c_{1} = 2$, $c_{2} = -6$, $c_{3} = 2$, $c_{4} = 3$. In this case potential (\ref{6}) looks as:

\begin{equation}
\label{12}
W^{(8)} = {\rm{Re}} \, \large[2 \, W_{1} - 6 \, W_{2} + 2 \, W_{3} + 3 \, W_{4}\large],
\end{equation}
and, according to (\ref{5}), (\ref{7}), (\ref{9}), (\ref{11}), it possesses the following stable minima in three special cases of the items 1-3 above:

\begin{equation}
\label{13}
\begin{array}{c}
\bar{s}_{12} = \sqrt{\frac{m_{d}}{|m_{s}|}} = 0.224 \pm 0.015; \qquad \bar{s}_{23} = \frac{|m_{s}|}{m_{b}} = (2.22 \pm 0.25) \cdot 10^{-2}; \\
\qquad \bar{s}_{13} = \frac{\sqrt{m_{d}\,|m_{s}|}}{m_{b}} = (4.9 \pm 0.9) \cdot 10^{-3}.
\end{array}
\end{equation}
It is seen from comparison of these calculated extremal values with experimental data (\ref{3}) that ${\bar{s}}_{12}$ coincides with experimental value of $s_{12}$, ${\bar{s}}_{13}$ is quite close to the observed $s_{13}$, whereas ${\bar{s}}_{23}$ is twice less than the experimental value of $s_{23}$ in (\ref{3}). This missing factor 2 in expression for $s_{23}$ has been a longstanding problem in the "texture zeroes" approach; in recent paper \cite{Fritzsch} (b) the corrections to this approach resolving this "factor 2" problem are proposed.

In the approach of the present paper it is also possible to resolve this "factor 2" problem with a certain choice of integer constants $c_{i}$ in potential (\ref{6}). Namely for $c_{1} = 7$, $c_{2} = - 24$, $c_{3} = 10$, $c_{4} = 9$, that is for potential (\ref{6})

\begin{equation}
\label{14}
W^{(8)} = {\rm{Re}} \, \large[7 \, W_{1} - 24 \, W_{2} + 10 \, W_{3} + 9 \, W_{4}\large],
\end{equation}
extremal values (\ref{5}), (\ref{9}), (\ref{11}) of sinuses of the $V_{CKM}$ angles 

\begin{equation}
\label{15}
\bar{s}_{12} = \sqrt{\frac{m_{d}}{|m_{s}|}}; \qquad \bar{s}_{23} = 2 \frac{|m_{s}|}{m_{b}}; \qquad \bar{s}_{13} = \frac{\sqrt{m_{d}\,|m_{s}|}}{m_{b}}
\end{equation}
coincide with the experimentally observed values (\ref{3}). 

It may be shown that results (\ref{9}), (\ref{15}) for $\bar{s}_{23}$ remain valid even if we consider total potential $W(s_{12}, s_{23}, s_{13})$ depending on three mixing angles. Unfortunately this is not true for results (\ref{5}), (\ref{11}) or (\ref{15}) for $\bar{s}_{12}$ and $\bar{s}_{13}$ since the crossing terms of full potential of type $s_{12}^{2} \, s_{23}^{2}$ etc. destroy these extrema.

Surely the choice (\ref{14}) of potential $W$ looks quite arbitrary, but perhaps no more arbitrary than corrections in the "texture zeroes" approach in \cite{Fritzsch} (b). And contrary to the flavor-invariant approach of the present paper, the drawback of the "texture zeroes" approach is its flavor noninvariance as it was clearly shown by Cecilia Jarlskog in \cite{Jarlskog} (b).

\section{Conclusion: Questions for Future}

\qquad Thus we considered Yukawa couplings as non-dynamical parameters and received phenomenologically viable values of $V_{CKM}$ mixing angles with extremizing of tree level potential of Yukawa matrices. The mathematical apparatus that makes it possible to obtain physically meaningful results within the framework of the "potential" flavor-invariant approach is, perhaps, the main result of this paper. 

The extremal points of potential $W$ may be considered as some self-consistency conditions imposed on the primordial mass matrices $\hat{M}_{u,d}$. To find the physical justification of the form of potential $W(\hat{M}_{u}, \hat{M}_{d})$ and of the corresponding self-consistency conditions is perhaps an interesting task for future. Results of this paper may hopefully stimulate this direction of thought.

Perhaps quantum vacuum diagrams containing $\hat{V}_{CKM}$ may provide the looked for potential $W$. The diagrams with the charged gauge boson internal Green functions have $\hat{V}_{CKM}$ in their every ${\bar{\psi}}_{ui}\psi_{dj}W_{\mu}^{\pm}$ vertex. The same is true for quantum vacuum diagrams with internal charged scalar field links. Thus the contributions of these diagrams to quantum potential will include flavor invariant traces of 2-nd, 4-th and higher powers in $\hat{V}_{CKM}$ similar to those used in this paper. Is it possible to build quantum potential which extremization provides sensible dependencies of mixing angles on mass ratios?

There are two serious drawbacks of the proposed approach.

Firstly. The extremization of potential $W$ (\ref{1}) is considered in the paper for three quite limited cases: (1) mixing of two first generations - $W(s_{12}, 0, 0)$; (2) mixing of two last generations - $W(0, s_{23}, 0)$; and mixing of first and third generations - $W(0, 0, s_{13})$. Extremization of the full potential $W^{(8)}(s_{12}, s_{23}, s_{13})$, as it was said above, does not change our result for $\bar{s}_{23}$ but destroys results for $\bar{s}_{12}$ and $\bar{s}_{13}$. Then the question arises: is it possible to justify the use of separate potentials for mixing angles of every couple from three generations?

Secondly. The potential $W(\hat{M}_{u}, \hat{M}_{d})$ permitting to obtain the meaningful dependence on quark mass ratios of the CP-violating phase $\delta$ (\ref{2}) or of the Jarlskog factor $J$ (\ref{4}) was not found here. Last line of general expression (\ref{20}) in the Appendix demonstrates the trivial linear dependence on $J$ of every 4-th power in $\hat{V}_{CKM}$ flavor invariant. Does it mean that interesting functional of $J$ may be received only with account of the higher powers in $\hat{V}_{CKM}$ flavor invariants which calculation is not a simple mathematical task?

\section*{Acknowledgements} Author is grateful to Valery Rubakov and Mikhail Vysotsky for inspiring consultations on the ABC of Standard Model and its problems.

\section*{Appendix}

\qquad Formulas below present the expressions of the alternating "chain" products of arbitrary diagonal 3x3 matrices $\hat{\alpha} = {\rm{diag}}(\alpha_{1}, \alpha_{2}, \alpha_{3})$ etc. with arbitrary 3x3 unitary matrices $\hat{V}$ and $\hat{V}^{\dag}$ through the differences of eigenvalues $\alpha_{12} = \alpha_{1} - \alpha_{2}$ etc., and moduli of the "angle" elements of $\hat{V}$:

\begin{equation}
\label{16}
p = 1 - |v_{11}|^{2}; \quad s = 1 - |v_{33}|^{2}; \quad q = |v_{13}|^{2}; \quad r = |v_{31}|^{2}. 
\end{equation}

For $\hat{V}_{CKM}$ (\ref{2}):

\begin{equation}
\label{17}
\begin{array}{c}
p = s_{12}^{2} + s_{13}^{2} - s_{12}^{2} \, s_{13}^{2}; \quad s = s_{23}^{2} + s_{13}^{2} - s_{23}^{2} \, s_{13}^{2}; \quad q = s_{13}^{2};   \\
r = s_{13}^{2} + s_{12}^{2} \, s_{23}^{2} - s_{12}^{2} \, s_{13}^{2} - s_{23}^{2} \, s_{13}^{2} + (s_{12}\, s_{23} \, s_{13})^{2} - \\ 
- 2 \, s_{12}\, c_{12} \, s_{23} \, c_{23} \, s_{13} \, \cos\delta.
\end{array}
\end{equation}

Thus for the "chain" trace with two CKM matrices it is obtained:

\begin{equation}
\label{18}
{\rm{Tr}}\{\hat{\alpha}\hat{V}\hat{\beta}\hat{V}^{\dag}\} = \sum_{i=1}^{3}\alpha_{i}\beta_{i} - [p \, \alpha_{12}\beta_{12} + q \, \alpha_{12}\beta_{23} + r \, \alpha_{23}\beta_{12} + s \, \alpha_{23}\beta_{23}].
\end{equation}

For three specific cases considered above we have with account of (\ref{17}):

\begin{equation}
\label{19}
\begin{array}{c}
1) \qquad s_{12} \ne 0: \,\, \, \,  {\rm{Tr}}\{\hat{\alpha}\hat{V}\hat{\beta}\hat{V}^{\dag}\} = \sum_{i=1}^{3}\alpha_{i}\beta_{i} - s_{12}^{2} \, \alpha_{12}\beta_{12}; \\
2) \qquad s_{23} \ne 0: \,\, \, \, {\rm{Tr}}\{\hat{\alpha}\hat{V}\hat{\beta}\hat{V}^{\dag}\} = \sum_{i=1}^{3}\alpha_{i}\beta_{i} - s_{23}^{2} \, \alpha_{23}\beta_{23}; \\
3) \qquad s_{13} \ne 0: \,\, \, \, {\rm{Tr}}\{\hat{\alpha}\hat{V}\hat{\beta}\hat{V}^{\dag}\} = \sum_{i=1}^{3}\alpha_{i}\beta_{i} - s_{13}^{2} \, \alpha_{13}\beta_{13}.
\end{array}
\end{equation}

The "chain" trace with four CKM matrices looks as:

\begin{equation}
\label{20}
\begin{array}{c}
{\rm{Tr}}\{\hat{\alpha}\hat{V}\hat{\beta}\hat{V}^{\dag}\hat{\gamma}\hat{V}\hat{\delta}\hat{V}^{\dag}\} = \sum_{i=1}^{3}\alpha_{i}\beta_{i}\gamma_{i}\delta_{i} + \\  
+ [p \, \alpha_{12}\beta_{12} + q \, \alpha_{12}\beta_{23} + r \, \alpha_{23}\beta_{12} + s \, \alpha_{23}\beta_{23}] \, \cdot \\
\cdot \, [p \, \gamma_{12}\delta_{12} + q \, \gamma_{12}\delta_{23} + r \, \gamma_{23}\delta_{12} + s \, \gamma_{23}\delta_{23}] + \\
+ \frac{1}{2} \, (q\,r - p\, s)\, (\alpha_{13}\gamma_{23} - \alpha_{23}\gamma_{13}) \, (\beta_{13}\delta_{23} - \beta_{23}\delta_{13}) - \\ 
- p \, [\alpha_{12}\gamma_{12}\beta_{12}\delta_{12} + (\alpha_{1}\gamma_{1} - \alpha_{2}\gamma_{2}) \, (\beta_{1}\delta_{1} - \beta_{2}\delta_{2})] - \\ 
- s \, [\alpha_{23}\gamma_{23}\beta_{23}\delta_{23} + (\alpha_{2}\gamma_{2} - \alpha_{3}\gamma_{3}) \, (\beta_{2}\delta_{2} - \beta_{3}\delta_{3})] + \\ 
+ q \, [\alpha_{12}\gamma_{12}\beta_{23}\delta_{23} + (\alpha_{2}\gamma_{2} - \alpha_{1}\gamma_{1}) \, (\beta_{2}\delta_{2} - \beta_{3}\delta_{3}) - \\ 
- \frac{1}{2} \, (\alpha_{12}\gamma_{13} + \alpha_{13}\gamma_{12}) \, (\beta_{13}\delta_{23} + \beta_{23}\delta_{13})] + \\ 
+ r \, [\alpha_{23}\gamma_{23}\beta_{12}\delta_{12} + (\alpha_{2}\gamma_{2} - \alpha_{3}\gamma_{3}) \, (\beta_{2}\delta_{2} - \beta_{1}\delta_{1}) - \\ 
- \frac{1}{2} \, (\alpha_{13}\gamma_{23} + \alpha_{23}\gamma_{13}) \, (\beta_{12}\delta_{13} + \beta_{13}\delta_{12})] + \\ 
+ i J \, (\alpha_{13}\gamma_{23} - \alpha_{23}\gamma_{13}) \, (\beta_{13}\delta_{23} - \beta_{23}\delta_{13}),
\end{array}
\end{equation}
where $J$ see in (\ref{4}). It is easy to show that for $\hat{\alpha} = \hat{m}_{u}$, $\hat{\beta} = \hat{m}_{d}$, $\hat{\gamma} = \hat{m}_{u}^{2}$, $\hat{\delta} = \hat{m}_{d}^{2}$ imagery part of trace (\ref{20}) given in the last line of (\ref{20}) is equal to half of Jarlskog determinant (\ref{4}), as it could be expected.

For three specific cases, with account of (\ref{17}), expression (\ref{20}) comes to:

1) $s_{12} \ne 0$, $s_{23} = s_{13} = 0$:

\begin{equation}
\label{21}
\begin{array}{c}
{\rm{Tr}}\{\hat{\alpha}\hat{V}\hat{\beta}\hat{V}^{\dag}\hat{\gamma}\hat{V}\hat{\delta}\hat{V}^{\dag}\} = \sum_{i=1}^{3}\alpha_{i}\beta_{i}\gamma_{i}\delta_{i} + s_{12}^{4} \, \alpha_{12}\gamma_{12}\beta_{12}\delta_{12} - \\  
- s_{12}^{2} \, [\alpha_{12}\gamma_{12}\beta_{12}\delta_{12} + (\alpha_{1}\gamma_{1} - \alpha_{2}\gamma_{2}) \, (\beta_{1}\delta_{1} - \beta_{2}\delta_{2})].
\end{array}
\end{equation}

2) $s_{23} \ne 0$, $s_{12} = s_{13} = 0$:

\begin{equation}
\label{22}
\begin{array}{c}
{\rm{Tr}}\{\hat{\alpha}\hat{V}\hat{\beta}\hat{V}^{\dag}\hat{\gamma}\hat{V}\hat{\delta}\hat{V}^{\dag}\} = \sum_{i=1}^{3}\alpha_{i}\beta_{i}\gamma_{i}\delta_{i} + s_{23}^{4} \, \alpha_{23}\gamma_{23}\beta_{23}\delta_{23} - \\  
- s_{23}^{2} \, [\alpha_{23}\gamma_{23}\beta_{23}\delta_{23} + (\alpha_{2}\gamma_{2} - \alpha_{3}\gamma_{3}) \, (\beta_{2}\delta_{2} - \beta_{3}\delta_{3})].
\end{array}
\end{equation}

3) $s_{13} \ne 0$, $s_{12} = s_{23} = 0$:

\begin{equation}
\label{23}
\begin{array}{c}
{\rm{Tr}}\{\hat{\alpha}\hat{V}\hat{\beta}\hat{V}^{\dag}\hat{\gamma}\hat{V}\hat{\delta}\hat{V}^{\dag}\} = \sum_{i=1}^{3}\alpha_{i}\beta_{i}\gamma_{i}\delta_{i} + s_{13}^{4} \, \alpha_{13}\gamma_{13}\beta_{13}\delta_{13} - \\  
- s_{13}^{2} \, [\alpha_{13}\gamma_{13}\beta_{13}\delta_{13} + (\alpha_{1}\gamma_{1} - \alpha_{3}\gamma_{3}) \, (\beta_{1}\delta_{1} - \beta_{3}\delta_{3})].
\end{array}
\end{equation}


\begin{thebibliography}{99}
\bibitem{Isidori}(a) G. Isidori, {\it{"Flavor Physics and Implication for New Phenomena"}}, Contribution to "The Standard Theory up to the Higgs discovery - 60 years of CERN", Ed. by L. Maiani and G. Rolandi, arXiv:1507.00867 [hep-ph]; 

(b) D.A. Faroughy, G. Isidori, F. Wilsch, K. Yamamoto, {\it{"Flavor symmetries in the SMEFT"}}, J. High Energ. Phys. 2020, 166 (2020), arXiv:2005.05366 [hep-ph].
\bibitem{Xing}Z.-Z. Xing, {\it{Flavor structures of charged fermions and massive neutrinos}}, Physics Reports 854 (2020) 1-147, arXiv:1909.09610 [hep-ph].
\bibitem{Feruglio}(a) F. Feruglio, {\it{"Pieces of the Flavor Puzzle"}}, Eur. Phys. J. C 75 (2015), no. 8 373, arXiv:1503.04071 [hep-ph]; 

(b) F. Feruglio, A. Romanino, {\it{"Lepton Flavor Symmetries"}}, Rev. Mod. Phys. 93, 015007 (2021), arXiv:1912.06028 [hep-ph].
\bibitem{Peskin}M.E. Peskin, D.V. Schroeder, {\it{An Introduction to Quantum Field Theory}}, Perseus Books Publ. L.L.C., Massachusets, 1995.
\bibitem{masses}G.-Y. Huang, S. Zhou, {\it{"Precise Values of Running Quark and Lepton Masses in the Standard Model"}}, Phys. Rev. D 103, 016010 (2021), arXiv:2009.04851 [hep-ph].
\bibitem{angles}P.A. Zyla et.al. [Particle Data Group], {\it{"Review of Particle Physics"}}, PTEP 2020 (2020) no. 8, 083C01.
\bibitem{Wolfenstein}L. Wolfenstein, Phys. Rev. Lett. 51 (1983) 1945.
\bibitem{Wilczek}F. Wilczek and A. Zee, {\it{“Discrete flavor symmetries and a formula for the Cabibbo angle”}}, Phys. Lett. 70B 418 (1977). 
\bibitem{Fritzsch}H. Fritzsch, (a) {\it{"Calculating the Cabibbo angle”}}, Phys. Lett. 70B 436 (1977); (b) {\it{"Flavor Mixing of Quarks and a New Texture"}}, Preprint arXiv:2103.10336 [hep-ph].
\bibitem{Cabibbo}N. Cabibbo and L. Maiani, {\it{"Dynamical Interrelation of Weak, Electromagnetic and Strong Interactions and the Value of $\theta$"}}, Phys. Lett, {\bf{28B}}, 2, 131-135; in {\it{Evolution of particle physics}}, ed. M. Conversi (Academic Press Inc., New York, 1970) 4;

R. Gatto, G. Sartori and M. Tonin, {\it{"Weak Self-masses, Cabibbo angle, and Broken $SU_{2} \times SU_{2}$"}}, Phys. Lett. {\bf28B}, 2, 128-130.
\bibitem{Alonso1}R. Alonso, M.B. Gavela, G. Isidori, L. Maiani, {\it{Neutrino Mixing and Masses from a Minimum
Principle}}, . High Energ. Phys. 2013, 187 (2013), arXiv:1306.5927 [hep-ph]; 

R. Alonso, {\it{Fermion masses and mixing from a minimum principle}}, Preprint arXiv:1405.5831 [hep-ph]; 

R. Alonso, M.B. Gavela, D. Hernandez, L. Merlo, and S. Rigolin, {\it{Leptonic Dynamical Yukawa Couplings}},  J. High Energ. Phys. 2013, 69 (2013), arXiv:1306.5922 [hep-ph].
\bibitem{Ambrosio}G. D’Ambrosio, G. F. Giudice, G. Isidori, and A. Strumia, {\it{Minimal Flavor Violation: an effective field theory approach}}, Nucl. Phys. {\bf{B645}} (2002) 155, arXiv:0207036 [hep-ph].
\bibitem{Harrison}P.F. Harrison and W.G. Scott, {\it{Covariant Extremization of Flavor-Symmetric Jarlskog Invariants and the Neutrino Mixing Matrix}}, Phys.Lett. B628 (2005) 93, arXiv:0508012 [hep-ph].
\bibitem{difficulties}B. Grinstein, M. Redi, and G. Villadoro, {\it{Low Scale Flavor Gauge Symmetries}}, JHEP 1011:067,2010, arXiv:1009.2049 [hep-ph].
\bibitem{Jarlskog}C. Jarlskog, (a) Zeit. f. Phys. C29 (1985) 491; Phys. Rev. Lett. 55 (1985) 1039; Phys. Rev. D35 (1987) 1685; ibid D36 2128; (b) {\it{On Invariants of Quark and Lepton Mass Matrices in the Standard Model}}, Comptes Rendus Physique, Volume 13, Issue 2, March 2012, Pages 111-114, arXiv:1102.2823 [hep-ph].
\bibitem{Branco}J. Bernabeu, G.C. Branco and M. Gronau, Phys. Lett. 169B, 243 (1986); 

I. Dunietz, O.W. Greenberg and D.-D. Wu, Phys. Rev. Lett. 55, 2935 (1985); 

J.P. Silva, {\it{Phenomenological aspects of CP violation}}, arXiv:0410351 [hep-ph].
\bibitem{invariants}E.E. Jenkins and A.V. Manohar, {\it{Algebraic Structure of Lepton and Quark Flavor Invariants and CP Violation}}, JHEP 0910 (2009) 094, arXiv:0907.4763 [hep-ph].
\bibitem{Bingrong1}Y. Wang, B. Yu and S. Zhou, {\it{"Flavor Invariants and Renormalization-group Equations in the Leptonic Sector with Massive Majorana Neutrinos"}}, arXiv:2107.06274 [hep-ph]; 

B. Yu and S. Zhou, {\it{Hilbert Series for Leptonic Flavor Invariants in the Minimal Seesaw Model}}, arXiv:2107.11928 [hep-ph].
\bibitem{Lu}J. Lu, {\it{"Comment on 'Flavor Invariants and Renormalization-group Equations in the Leptonic Sector
with Massive Majorana Neutrinos'”}}, JHEP 02 (2022) 135, arXiv:2111.02729 [hep-ph].
\bibitem{Bingrong2}B. Yu and S. Zhou, {\it{"Spelling Out Leptonic CP Violation in the Language of Invariant Theory"}}, arXiv:2203.00574 [hep-ph]; {\it{"CP violation and flavor invariants in the seesaw effective field theory"}}, arXiv:2203.10121 [hep-ph].
\end{thebibliography}
\end{document}